\documentclass[reprint,aps,pre]{revtex4-2}

\usepackage{dcolumn}
\usepackage{bm}

\usepackage[utf8]{inputenc}
\usepackage[T1]{fontenc}
\usepackage{mathptmx}
\usepackage{etoolbox}
\usepackage{amsmath}

\usepackage{graphicx}    
\usepackage{siunitx}
\usepackage[dvipsnames]{xcolor}
\usepackage{float} 
\usepackage{tabularx}
\usepackage{amssymb}
\usepackage{ulem}
\usepackage{multirow}

\usepackage[version=3]{mhchem} 
\usepackage{chemformula} 

\usepackage{subcaption}
\usepackage{diagbox}

\newcommand{\corr}[1]{#1}

\begin{document}

\title{Response of fluorescent molecular rotors in ternary macromolecular mixtures}
  
\author{Mingshan Chi}
\affiliation{Laboratoire du Futur, (LOF) - Syensqo - CNRS - Universite de Bordeaux, UMR 5258, Bordeaux, 33600 Pessac, France}
\author{Anh-Thy Bui}
\affiliation{Univ. Bordeaux, CNRS, Institut des Sciences Mol{\' e}culaires (UMR5255), 351 cours de la Libération, 33405 Talence cedex, France}
\author{Pierre Lidon}
\email{pierre.lidon@u-bordeaux.fr}
\author{Yaocihuatl Medina Gonzalez}
\email{yaocihuatl.medina-gonzalez@u-bordeaux.fr}
\affiliation{Laboratoire du Futur, (LOF) - Syensqo - CNRS - Universite de Bordeaux, UMR 5258, Bordeaux, 33600 Pessac, France}

\date{\today}

\begin{abstract}
  For a few decades, Fluorescent Molecular Rotors have been commonly employed as local probes of microviscosity in complex materials. However, without proper calibration, relating microviscosity to a physical parameter is unclear, which strongly limits their quantitative use in biological media for instance. In this study, the response of a molecular rotor in binary and ternary macromolecular aqueous solutions of polyethylene glycol (PEG) of different molecular weights is investigated in order to better rationalize the sensitivity of rotors to their cybotactic environment. More precisely, for the investigated composition range of ternary mixtures, it is shown that a linear mixing rule applies for fluorescence lifetime with the proportion of the two PEG, and with an increasing ratio of heavy PEG leading to larger lifetimes. These results allow to test more precisely the free volume theory, which has been proposed in the context of probing glass transition. Analysis show that while this theory semi-quantitatively captures the observation, its precise use raises some questions.
\end{abstract}

\maketitle

\section{Introduction}

Fluorescent molecular rotors (FMRs) are fluorescent molecules displaying two competitive relaxation pathways after photoexcitation. Aside of the conventional fluorescent path, they can return to the ground state through a non-radiative process, involving intramolecular twisting motion. As such rotation is hindered by the friction with the surrounding environment, radiative relaxation is favored in more viscous environments, leading to more intense fluorescence. FMRs thus appear as viscosity-sensitive fluorescent probes, which have been successfully applied in numerous systems such as simple liquids in different conditions of temperature, pressure and confinement~\cite{hosny_2013,garg_2020,polita_2020,caporaletti_2022}, polymer films and solutions~\cite{meyer_1990,zhu_2007,lee_2011b,kim_2023e}, ionic liquids~\cite{paul_2008,singh_2016} or biological fluids~\cite{kung_1989,akers_2005,daus_2023}, with possibility to monitor evolution of polymerizing or gelating systems for instance~\cite{schmitt_2022,raeburn_2015,ledizescastell_2024a}. Recent developments were devoted to the grafting of FMRs on optical fibers~\cite{haidekker_2006,lichlyter_2009} or polymers~\cite{jin_2017,nalatamby_2025} to formulate viscosity-sensitive materials, that can be used in situ. Finally, the coupling of FMRs with Fluorescence Lifetime Imaging Microscopy (FLIM) opened perspectives for viscosity mapping in heterogeneous systems, like emulsions~\cite{velandia_2023,bittermann_2023}, lipid membranes and vesicles~\cite{nipper_2008,wu_2013,dent_2015,vysniauskas_2016,adrien_2022} or biological cells and tissues~\cite{li_2016,zhang_2019b,michels_2020,paezperez_2024,xie_2025b}. 

In all these applications, FMRs were mostly used as qualitative contrast agents. In order to obtain quantitative information, the response of fluorescence of FMRs first needs to be calibrated against a physical parameter. Fluorescence quantum yield $\varphi$ is usually related to viscosity $\eta$ by a powerlaw, known as F{\"o}rster-Hoffmann equation~\cite{forster_1971}:
\begin{equation}
\log \phi = x \log \eta + C,
\label{eq:FH_QY}
\end{equation}
\noindent in which $x$ is a constant characterizing FMR sensitivity and $C$ is an instrumental constant. Being proportional to the quantum yield, fluorescence intensity $I$ and lifetime $\tau$, which can both be directly measured, also follow such powerlaw evolution:
\begin{equation}
\left\{ \begin{array}{l}
\log I = x \log \eta + C_I \\
\log \tau = x \log \eta + C_\tau
\end{array} \right. .
\label{eq:FH_I_tau}
\end{equation}
\noindent After calibration with homogeneous solutions of known viscosity to determine values of parameters $x$ and $C$, local viscosity can thus be retrieved from standard fluorescence measurements~\cite{bunton_2014,nalatamby_2023,gibouin_2024,mirzahossein_2025}. However, such equation is generally valid only in a limited range of viscosities and parameters $x$ and $C$ are not universal, depending both on the chemical nature of the FMR and of its environment. The calibration should therefore be performed in a material sufficiently similar to these to be investigated, which strongly limits the possibility to interpret FMRs fluorescence in more complex systems, such as biological materials.

A more precise understanding of the sensitivity of a FMR to its environment is thus needed. Theoretically, FMRs can be treated as an elementary molecular machine, in which microscopic motion is powered by light absorption~\cite{klok_2009,roy_2021b,roy_2024a}: efficiency of such process is limited by the interactions of the moving parts of the molecule with its cybotactic environment, which is usually described by the empirical concept of ``microviscosity'', ``fluidity'', or ``microfriction'' ~\cite{straube_2020,kiefer_2025} \corr{(and sometimes called ''nanoviscosity'' to make a difference with the mesoscopic viscous force undergone by nano- or micrometric probes in microrheology)}. F{\"o}rster-Hoffmann equation can then be theoretically justified by modeling the rotor as a thermally-excited harmonic oscillator damped by the local microviscosity~\cite{forster_1971,rothenberger_1983,suja_2025}, yet this does not clarify the relationship between this microscopic parameter and the hydrodynamic viscosity, quantifying resistance to flow and defined at a larger, mesoscopic length scale. More generally, this question is related to the general issue of the prediction of chemical transformation rates. In liquid state, these occur through short-range interactions and random collisions in the immediate environment of the involved molecules, driven by Brownian motion. At macroscopic scale, they can thus be described as activated processes, following the empirical Arrhenius law: statistical thermodynamics basis of such equation have been laid by Eyring's transition state theory, giving a clear interpretation of the activation energy.  \corr{Beyond transition state theory, progress have been made by incorporating memory effects of microviscosity on microscopic transformation rates and showing that zero-frequency viscosity is not sufficient to describe macroscopic dynamics, but relaxation spectrum of the surrounding medium should also be accounted for.~\cite{northrup_1980,grote_1980} In particular, it has recently been shown that deviations from Stokes-Einstein and Kramers equations are observed in the conformational isomerization rate of large molecules.~\cite{dalton_2024} However, a complete description of reaction rates} would require the development of a complete kinetic theory in liquid state which is still out of reach.

To investigate this question, macromolecular systems represent an appropriate model since their viscosity results from interactions of objects at mesoscopic scale, larger than these expected to influence the motion of FMRs, \corr{and their microscopic relaxation spectrum is expected to be more complex than for molecular fluids}. Recently, study of the response of a specific FMR in semi-dilute aqueous solution of polyethylene glycols (PEG) of different molecular weights was performed by Bittermann et al.~\cite{bittermann_2021a}. They clearly showed that even in such solutions of very similar chemical nature, the FMR fluorescence was not solely determined by viscosity, but was also influenced by the PEG molecular weight, and rescaled all their results by using the polymer weight concentration. To explain these observations and connect them to the F{\"o}rster-Hoffmann equation, they proposed that the fluorescent response is related to the diffusion length, which is set by the characteristic blob size of the semi-dilute polymer solution~\cite{kohli_2012}. However, the concept of blob is limited to the context of polymer solutions and their theory thus cannot be extended to other systems. Moreover, in binary systems, the weight fraction appears as the only quantity that is descriptive of the composition at microscopic scale, making it difficult to decipher the effect of different parameters on microviscosity.

In this article, the fluorescent response of a FMR designed to be operative at the low viscosity range is investigated in macromolecular aqueous solutions of PEG. Viscosity and density of binary and ternary mixtures of water with one or two PEGs of different molecular weights were measured to detect correlations with fluorescence lifetime, and thus microviscosity. In binary solutions, the obtained results were overall consistent with these reported by Bittermann et al.: they showed that the F{\"o}rster-Hoffmann law is valid in the different solutions, yet with an exponent depending on the molecular weight. The collapse of data when plotted as a function of the mass fraction of polymer is however less satisfactory. More interestingly in ternary solutions, it was observed that, for a given PEG couple and at fixed overall polymer weight fraction, lifetime evolves linearly with the proportion of the two PEG, higher lifetime (and hence microviscosity) being obtained with higher proportion of heavy PEG.

To explain these observations, results are analyzed in terms of free volume theory, a semi-empirical concept stemming from the theory of glass transition in polymer blends,~\cite{fox_1950} which has been suggested to be a good descriptor for microviscosity~\cite{loutfy_1982}. It is shown that experimental results are qualitatively consistent with a free volume approach, yet a fully quantitative comparison is difficult: in particular, computing free volume requires the estimation of a ``zero temperature volume'' which is hard to obtain in aqueous solutions. Two approaches are proposed, either considering free volume as a global or local parameter, which satisfactorily describe the measurements. These results thus deepen our understanding of FMR response in liquids. Besides, this study provides some systematic measurements of the viscosity and density of ternary solutions, together with appropriate mixing rules. Such results are interesting for applied science as PEG polymers are inexpensive, easy to formulate, and commonly employed in industry.

\section{Materials and methods}

\subsection{Rotor synthesis}

\begin{figure}[!htb]
\includegraphics{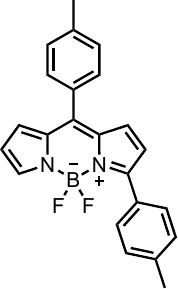}
\caption{Structure of the molecular rotor used in this study. \label{fig:rotor_structure}}
\end{figure}

Chemicals were obtained from commercial sources (Sigma-Aldrich, Fisher Sci., DougDiscovery, BLDPharm, TCI) and used without further purification. The rotor (Fig.~\ref{fig:rotor_structure}) was synthesized by Suzuki cross-coupling from its 3-brominated precursor according to reported procedures~\cite{rao_2010,bahaidarah_2014} and its purity attested by $^1$H and $^{13}$C NMR analyzes performed in deuterated chloroform on Bruker Avance I 300 MHz spectrometers that are available at the CESAMO platform of ISM (see Supp. Inf. for description of spectra).

Molecular rotors were stored in the dark in amber vials to avoid long term photobleaching, and it was verified that no significant bleaching occurred over the duration of experiments.

\subsection{Mixture formulation and characterization}

In this study, different binary and ternary aqueous solutions of ethylene glycol or PEG were prepared. Pure PEG of different molecular weight were used without further purification (Merck, CAS: 107-21-1 (EG) and 25322-68-3(PEG)). For clarity, in what follows, ethylene glycol is referred to as PEG of molecular weight $M=\SI{62}{\gram\per\mole}$. Their mass distributions were characterized at $\SI{26}{\celsius}$ by SEC-MALS analysis (HPLC Ultimate 3000 from ThermoFisher Scientific, coupled with MALES HELEOS II and Optilab T-rEX from Wyatt Technology) at the POLYCAR characterization platform. The polydispersity was around $\SI{10}{\percent}$ for PEG-400 sample, and below $\SI{5}{\percent}$ for samples of higher molecular weight. For all samples, the average molecular weights were consistent with their nominal values indicated by the manufacturer within a $\SI{10}{\percent}$ discrepancy. In the following, the PEG are designated by their nominal values, but measured number averaged molecular weights are used for calculations. PEG solutions are prepared by mixing PEG with deionized water, with continuous stirring until reaching a homogeneous solution. Several hours were necessary to solubilize PEG of high molecular weights. After formulation, solutions were stored in closed vials at room temperature and used in the following week to avoid evaporation. The list of investigated solutions is gathered in Table~S1 of Supp. Inf.

Density of the different solutions were measured with a densimeter (Anton Paar, DMA 4500M) at controlled temperature $\SI{20}{\celsius}$. Viscosity were measured using a stress-controlled rheometer (Netzsch Kinexus) in a Couette geometry (inner radius $\SI{27}{\milli\meter}$ and gap $\SI{1}{\milli\meter}$) at room temperature, by performing ramps of shear rates. All solutions were Newtonian, and reported viscosities were averaged over the range of shear rates.

Finally, FMR-PEG solutions were prepared by mixing a stock solution of rotor in ethanol with the appropriate mass of PEG solution, in order to reach a rotor concentration of $c_0 = \SI{1.8e-6}{\mole\per\liter}$ in all solutions. No aggregation-induced quenching was observed at this concentration, but fluorescence signal was intense enough to perform measurements.

\subsection{Fluorescence characterization}

Absorption and emission spectra of FMR-PEG solutions were acquired respectively using a UV-visible spectrophotometer (Agilent Technologies) and a fluorescence spectrophotometer (Cary Eclipse, Agilent Technologies).

Fluorescent lifetimes were measured using a frequency-domain FLIM (Lambert Instrument FLIM Attachment), whose principle is \corr{summarized in Supp. Inf. and more details can for instance be found in Ref.~\cite{lakowicz}}. The device was mounted on an inverted microscope (Olympus IX71) with a $10 \times/\mathrm{NA} = 0.30$ objective lens and dichroic mirror to separate excitation and emitted lights. Excitation light source was a light-emitting-diode of wavelength $\SI{520}{\nano\meter}$ modulated at a frequency of $\SI{40}{\mega\hertz}$. Emitted light was filtered by a long pass filter (Olympus, $530\times - \SI{560}{\nano\meter}$) and analyzed by a temperature-controlled CCD camera with a $504 \times \SI{512}{px^2}$, with acquisition over 12 different phases and adapted exposure time to obtain optimum signal without saturation. Samples consisted in droplets of the solution to be analyzed deposited on microscope glass slides, and reference was taken with similar droplets of Rhodamine 6G (CAS: 989-38-8), with tabulated lifetime of $\SI{4.08}{\nano\second}$. Lifetimes were retrieved using the commercial FLIM software and reported values are averaged over the whole field of view. Uncertainty analysis on such measurements are not straightforward, yet the resolution is of order of $\SI{0.1}{\nano\second}$.

\section{Results}

\subsection{Characterization of rotor fluorescence}

\begin{figure*}[!htb]
    \includegraphics{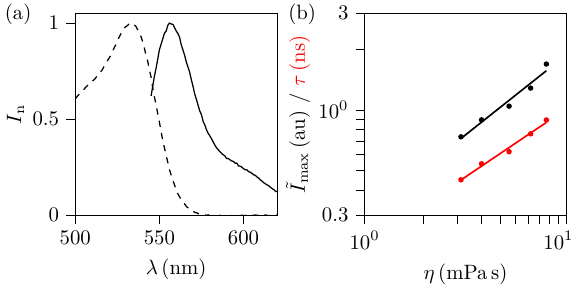}
    \caption{Fluorescence properties of rotor in ethylene glycol/water mixtures. (a) Absorption and emission spectra in pure ethylene glycol. Dashed line : absorption spectrum, normalized by maximum value and corrected for baseline, maximum of absorption is at $\lambda_\text{abs} = \SI{532}{\nano\meter}$; Continuous line : emission spectrum with excitation at maximum of absorption, normalized by maximum value, maximum of emission is at $\lambda_\text{em} = \SI{556}{\nano\meter}$. (b) Evolution of rotor normalized fluorescence intensity (black) and lifetime (red) in ethylene glycol/water mixtures (ethylene glycol weight fraction $w$ from $0.6$ to $1$). Continuous line are fit with Förster-Hoffmann equation~\eqref{eq:FH_I_tau} with $x=0.76$ for intensity and $x=0.65$ for lifetime.}
    \label{fig:carac_EG}
\end{figure*}

Among the diversity of reported FMR structures, BODIPY derivatives have emerged as a popular family of microviscosity probes owing to their low sensitivity to solvent polarity and to their high photostability. In the framework of this study, a 3-substituted BODIPY was selected for its enhanced response to the viscosity of the solvent (ascribed to a larger active sensing area than in the parent unsubstituted compound) and wider linearity range of the Förster-Hoffmann equation at low viscosities.~\cite{bahaidarah_2014} In addition to these advantageous features, the spectral properties of this FMR, adjusted by the extension of the $\pi$-conjugated system, make it optimal for use with our fluorescence microscopy setup ($\lambda_\text{exc} = \SI{520}{\nano\meter}$). Absorption and emission spectra in pure PEG-62 are reported in Fig.~\ref{fig:carac_EG}(a). The UV-visible absorption spectrum displays a maximum at $\lambda_\text{abs} = \SI{532}{\nano\meter}$, and upon photoexcitation at this wavelength, the emission spectrum displays a maximum at $\lambda_\text{em} = \SI{556}{\nano\meter}$.

The behavior of the rotor was first characterized in binary mixtures of water and PEG-62, with PEG-62 mass fractions between $w=0.6$ and $1$.
For the different mixtures, no significant deformation of the spectrum was observed nor a shift in the wavelength of maximum absorption and emission, and only an overall change in intensity with viscosity was obtained. The absorption and emission maxima (respectively, $A_\text{max}$ and $I_\text{max}$) were measured for the different solutions. In order to retrieve the parameters of the F{\"o}rster-Hoffmann equation~\eqref{eq:FH_I_tau}, maximum intensity should be used with care as it is not only proportional to the quantum yield $\varphi$ but also to the rotor concentration $c_0$. Solutions of similar concentration in rotor were prepared, and the normalized maximum intensities $\tilde{I}_\text{max} = I_\text{max} / A_\text{max}$ were systematically used to minimize possible residual effects related to rotor concentration. The evolution of this parameter with solution viscosity is reported in logarithmic coordinates in Fig.~\ref{fig:carac_EG}(b) (black dots) and follows a power-law with exponent $x = 0.76$. \corr{Fluorescence lifetimes $\tau$ of FMR in these solutions were also measured, and plotted against viscosity in Fig.~\ref{fig:carac_EG}(b) (red dots): again, a power-law evolution of close exponent $x = 0.65$ was observed. The obtained exponents are consistent with those reported in the literature for other BODIPY-based FMRs~\cite{kuimova_2008,nalatamby_2023}. As the lifetime is directly proportional to the quantum yield \corr{via $\phi = k_\text{r}\tau$ (in which $k_\text{r}$ is the radiative constant of the FMR, which remains unperturbed\cite{bahaidarah_2014})}, these results confirm the validity of F{\"o}rster-Hoffmann equation for the range of compositions investigated here. 

Such an equation should however be considered with care as the range of variation of the different parameters is below one decade. Considering the limited range of viscosity used to estimate these exponents, their values seem similar within experimental uncertainty. For a mono-exponential decay, both fluorescence intensity and lifetime are proportional to quantum yield and are supposed to follow F{\"o}rster-Hoffmann equation with the same exponent. In addition, polar plot analysis~\cite{leray_2012} and some additional measurements performed with time-domain lifetime measurements (data not shown, see SI for technical details) did not show any clear sign of multi-exponential decay.}

\subsection{Binary mixtures}

\begin{figure*}[!htb]
    \includegraphics{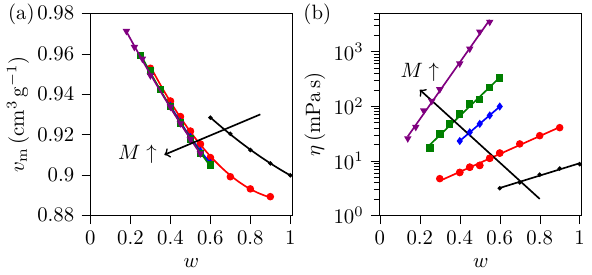}
    \caption{Evolution of properties of binary PEG/water mixture as a function of PEG mass fraction $w$, for different PEG molar weight ({\tiny $ ^\bullet$} PEG-62; {\color{red} $\bullet$} PEG-400; {\tiny \color{blue} $^\blacklozenge$} PEG-2000; {\tiny \color{OliveGreen} $\blacksquare$} PEG-6000; {\tiny \color{violet} $\blacktriangledown$} PEG-20000). Arrows indicate increasing molecular weight $M$. (a) Specific volume $v_\text{m}  = 1/\rho$. (b) Viscosity $\eta$. }
    \label{fig:carac_binary}
\end{figure*}

In this section, properties of binary mixtures of PEG with water are first considered. The specific volumes $v_\text{m} = 1/\rho$ and viscosity $\eta$ of the different solutions are reported as a function of the PEG mass fraction $w$ on Fig.~\ref{fig:carac_binary}(a) and (b) respectively.

On the investigated concentration range, the specific volume of solutions overall decreases with the fraction of PEG and with its molecular weight. Specific volumes of the different PEG are rather similar and overall follow an ideal mixing rule, except for PEG-62 solutions which display a higher specific volume, and for PEG-400 which displays deviations to ideality at the highest mass fractions. The curves can all be fitted with a second order polynomial:
\begin{equation}
v_\text{m} (M,w) = \alpha_v(M)w^2 + \beta_v(M) w + \gamma_v(M).
\label{eq:vm_binary}
\end{equation}
\noindent Evolutions of fitting coefficients $\alpha$, $\beta$ and $\gamma$ are presented in in Table~S2 of Supp. Inf.

Viscosity overall increases with the PEG fraction and molecular weight, and is well described by an exponential evolution:
\begin{equation}
\ln \eta(w,M) = (1-w) \ln \eta_0(M) + w \ln \eta_1(M),
\label{eq:eta_binary}
\end{equation}
\noindent in which $\eta_0$ and $\eta_1$ respectively correspond to extrapolated viscosity of pure water and pure PEG, in infinite dilution limit. Values of $\eta_0$ and $\eta_1$ are gathered in Table~S2 of Supp. Inf.

\begin{figure*}[!htb]
    \includegraphics{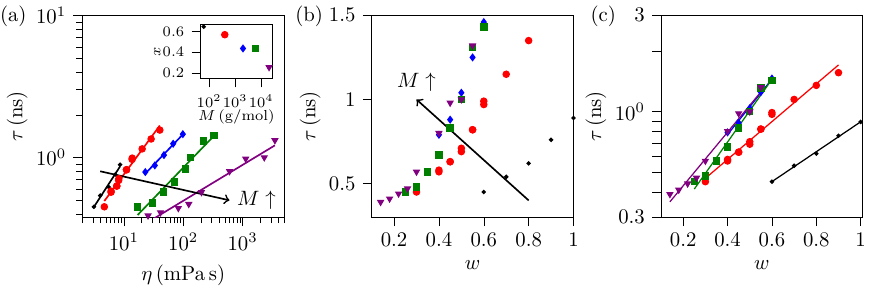}
    \caption{Evolution of rotor fluorescence lifetime $\tau$ in binary PEG/water mixtures, for different PEG molar weight ({\tiny $ ^\bullet$} PEG-62; {\color{red} $\bullet$} PEG-400; {\tiny \color{blue} $^\blacklozenge$} PEG-2000; {\tiny \color{OliveGreen} $\blacksquare$} PEG-6000; {\tiny \color{violet} $\blacktriangledown$} PEG-20000). Arrows indicate increasing molecular weight $M$. (a) Förster-Hoffmann representation of evolution of lifetime $\tau$ with viscosity $\eta$. Solid lines are fit with Equation~\eqref{eq:FH_I_tau} and inset represents evolution of exponent with PEG molar weight $M$. (b,c) Evolution of fluorescence lifetime $\tau$ of the FMR with PEG mass fraction $w$ in regular and semilogarithmic coordinates. }
    \label{fig:FH_binary}
\end{figure*}

\corr{As can be seen on Fig.~\ref{fig:FH_binary}(a), F{\"o}rster-Hoffman equation relating fluorescence lifetime and viscosity is verified, yet with an exponent $x$ depending on the molecular weight of the PEG consistently with results by Bittermann et al.~\cite{bittermann_2021a}.} However, contrary to what they observed, it can be seen on Fig.~\ref{fig:FH_binary}(b) that lifetime data do not collapse on a single mastercurve when plotted as a function of the mass fraction. Solutions with PEG of the largest molecular weights are comparable yet with observable discrepancies, but solutions of PEG-62 and PEG-400 lead to notably smaller lifetimes. Finally, consistently with Eq.~\eqref{eq:eta_binary}, logarithm of lifetime can be fitted linearly as a function of mass fraction $w$ along:
\begin{equation}
\ln \tau(w,M) = (1-w) \ln \tau_0(M) + w \ln \tau_1(M),
\label{eq:tau_binary}
\end{equation}
\noindent in which $\tau_0$ and $\tau_1$ respectively correspond to extrapolated lifetimes of the FMR in pure water and in pure PEG,  at infinite dilution limit. Values of $\tau_0$ and $\tau_1$ are gathered in Table~S2 of Supp. Inf.

\subsection{Ternary mixtures}
\label{sec:res_ternary}

In this section, properties of ternary mixtures of water and two distinct PEG are reported, with the two PEGs generically designated as PEG 1 and PEG 2. Denoting by $m_0$, $m_1$ and $m_2$ the respective mass of water, of PEG 1 and of PEG 2 (the list of investigated solutions is supplied in Table~S1 of Supp. Inf.), the composition of the solutions are described through the total PEG mass fraction, $w = (m_1 + m_2) / (m_0 + m_1 + m_2)$, and the proportion of PEG $2$, $W = m_2 / (m_1 + m_2)$. In this section, results are illustrated for mixtures of PEG-400 (taken as PEG 1) and PEG-2000 (taken as PEG 2), with mass fraction $w$ varying between $0.4$ and $0.6$ and proportion $W$ in PEG-2000 regularly distributed between $0.1$ and $0.9$. Similar results were obtained for the other investigated mixtures and are reported in Figures~S2--5 of Supp. Inf.

\begin{figure*}[!htb]
    \includegraphics{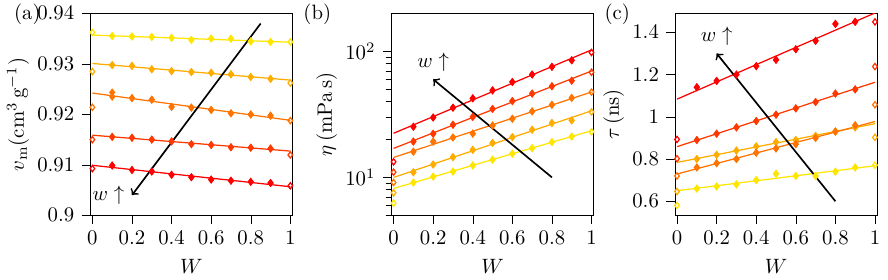}
    \caption{Properties of ternary mixtures of water, PEG-400 and PEG-2000 as a function of the proportion of PEG-2000, $W$. The color gradient represents the evolution of the total PEG mass fraction $w$, varying between $w=0.4$ and $0.6$ by step of $0.05$. The open symbols on axis $W=0$ and $W=1$ represent data obtained in binary PEG/water mixtures, displayed on Fig.~\ref{fig:carac_binary}. Continuous lines are linear fit along Eq.~\eqref{eq:ideal_ternary}. Arrows indicate increasing PEG mass fraction $w$. \corr{(a) Evolution of specific volume $v_\text{m}$. (b) Evolution of viscosity $\eta$ in semilogarithmic coordinates.} (c) Evolution of the fluorescence lifetime $\tau$ of the FMR.}
    \label{fig:example_ternary}
\end{figure*}

Overall, as displayed on Fig.~\ref{fig:example_ternary}, specific volume $v_\text{m}$, logarithm of the viscosity $\eta$ and lifetime $\tau$ evolved linearly with the proportion of heavier PEG, with decrease of specific volume, and increase of lifetime and viscosity. This first result is surprising, as for a given mass fraction of polymer, changing the length of the polymer chain is not expected to modify significantly the microviscosity of the immediate environment of the rotors. Also, this confirms that the local polymer fraction $w$ is not sufficient to predict rotor behavior.

Such behavior is observed for all the investigated mixture and suggests an ideal mixing behavior. For all ternary mixtures, the evolutions of the different parameters with the mass fraction $W$ of PEG 2 can thus be described by the following linear equations:
\begin{widetext}
\begin{equation}
\left\{\begin{array}{lll}
v_\text{m} (M_1,M_2,w,W) &= v^{(1)}_\text{m} (M_1,M_2,w) (1-W) &+ v^{(2)}_\text{m} (M_1,M_2,w) W \\
\ln \eta (M_1,M_2,w,W) &= \ln \eta^{(1)} (M_1,M_2,w) (1-W) &+ \ln \eta^{(2)} (M_1,M_2,w) W \\
\tau (M_1,M_2,w,W) &= \tau^{(1)} (M_1,M_2,w) (1-W) &+ \tau^{(2)} (M_1,M_2,w) W
\end{array} \right. .
\label{eq:ideal_ternary}
\end{equation}
\end{widetext}
\noindent in which $v^{(1/2)}_\text{m}$, $\ln \eta^{(1/2)}$ and $\tau^{(1/2)}$ are fitting parameters, determined for all the ternary mixtures.

These equations can be interpreted in the spirit of the thermodynamics of ideally dilute mixtures. The parameters $v^{(1)}_\text{m}$, $\ln \eta^{(1)}$ and $\tau^{(1)}$ (respectively $v^{(2)}_\text{m}$, $\ln \eta^{(2)}$ and $\tau^{(2)}$) are obtained by extrapolating linearly the measured properties to the limit $W = 0$ (respectively $W=1$).  For an ideal behavior on the whole composition range, they should coincide with the properties of binary solutions of water with PEG $1$ of PEG $2$ (displayed with open symbols on Fig.~\ref{fig:example_ternary}), which is not the case here. 

\corr{This notable difference between the properties of the limit solutions and the binary solutions} may seem surprising. A more detailed study has been performed for this mixture of PEG-400 and PEG-2000, for smaller fractions of PEG-2000 : the results are displayed in Figure~S11 of Supp. Inf. and evidence a non-ideal behavior in this regime, with a continuous transition towards the properties of binary solutions. The fitting parameters with superscript $(1)$ (respectively $(2)$) should thus be considered as properties of hypothetical, limit binary \corr{aqueous} solutions of PEG 1 (respectively 2) \corr{only at mass fraction $w$}, but with effective interactions accounting for the effect of the other PEG.

\begin{figure*}[!htb]
    \includegraphics{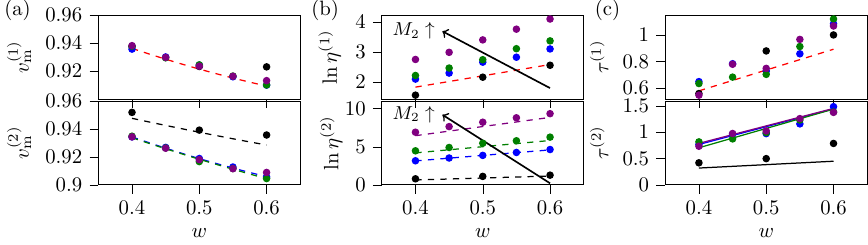}
    \caption{Evolution of fitting parameters in Eq.~\eqref{eq:ideal_ternary}, corresponding to limit binary solutions $(1)$ (top row) and $(2)$ (bottom row), as a function of total PEG mass fraction $w$ and for mixtures of PEG-400 (designated as PEG 1) with another PEG 2 ({$\bullet$} PEG-62, {\color{blue} $\bullet$} PEG-2000, {\color{OliveGreen} $\bullet$} PEG-6000, {\color{violet} $\bullet$} PEG-20000).  Dashed line corresponds to properties of binary solutions of water and PEG, obtained with Eqs.~\eqref{eq:vm_binary},~\eqref{eq:eta_binary} and~\eqref{eq:tau_binary}. Arrows indicate increasing molecular weight of PEG 2. (a) Specific volumes $v_\text{m}$, in $\SI{}{\centi\meter\cubed\per\gram}$. (b) Logarithm of viscosity $\ln \eta$ in $\SI{}{\milli\pascal\second}$. (c) Fluorescence lifetime $\tau$ of the FMR, in $\SI{}{\nano\second}$.}
    \label{fig:parameters_ternary}
\end{figure*}

These parameters are represented as a function of the overall PEG fraction $w$ on Fig.~\ref{fig:parameters_ternary}, for the mixtures involving PEG-400 designated as PEG 1. Results for other PEGs are displayed in Tables~S3--7 Figure~S6--10 of Supp. Inf. Overall, a linear evolution can be observed. In the case of specific volume, variations are moderate. For both limit solutions, data are consistent with the equivalent binary solutions (displayed as dashed line) and properties of limit solution $(1)$ are independent on the molecular weight of PEG 2: this agrees with a fully ideal behavior. For viscosity and lifetime however, clear deviations from ideality can be observed as the properties of the two limit solutions depend on the molecular weight of the other PEG and do not match these of binary solutions. 

\section{Discussion}

\subsection{Reminder of free volume theory for molecular rotors}

The experimental results reported above clearly show that neither the viscosity, nor the average mass fraction of PEG alone can describe the response of the FMR. \corr{This is consistent with the micro-environment of molecular rotors in the investigated macromolecular solutions, which is likely to be associated with  complex viscous relaxation spectrum.~\cite{dalton_2024} Experimentally, a full characterization of such phenomenon is out of reach, and for practical use, the design of more phenomenological approaches remains of interest.} 

Some early studies on photoisomerizable systems~\cite{gegiou_1968} and on FMRs~\cite{loutfy_1981,loutfy_1982} suggested that their dynamics was actually related to a semi-empirical concept, the free volume. This quantity was initially introduced to describe the glass transition of polymer blends~\cite{fox_1950,williams_1955} and the viscosity of liquids~\cite{doolittle_1951a,doolittle_1951b,doolittle_1951c,doolittle_1952a,doolittle_1952b,doolittle_1957}, and later mostly used~\cite{kovacs_1966,haward_1970} and refined~\cite{vrentas_1977a,vrentas_1977b} in the context of physico-chemistry of polymer melts and solutions. The coupling of Eyring's and free volume theories however captures semi-quantitatively many observations in a wide range of systems involving thermally activated processes~\cite{hao_2024}.

Several definitions of free volume can be found in the literature, but the one relevant for FMRs is the total free volume $v_\text{f}$~\cite{haward_1970,kim_2023e} corresponding to the difference between the total volume of the system, and the hard-sphere, incompressible volume of matter contained in it. More precisely, microscopic motion is dictated by the free volume fraction $\nu$ defined as:
\begin{equation}
\nu = \frac{v_\text{f}}{v_0} = \frac{v_\text{m} -v_0}{v_0}
\end{equation}
\noindent in which $v_\text{m}$ is the specific volume and $v_0$ the incompressible volume of matter per unit mass. As displacement at the microscopic scale requires molecules to fit inside available volumes, Doolittle proposed to describe viscosity $\eta$ as a function of free volume fraction~\cite{doolittle_1951b}:
\begin{equation}
\eta = A e^{B/\nu} 
\end{equation}

\noindent in which $A$ and $B$ are constant, depending on the chemical nature of the solution. Loutfy later suggested that rotation of a FMR requires some free space to rotate, proposing that fluorescence quantum yield is given by~\cite{loutfy_1982}:
\begin{equation}
\varphi = \varphi_0 e^{\tilde{x}/\nu}
\end{equation}
\noindent in which $\varphi_0$ is the quantum yield the FMR would have if intramolecular rotation was fully free, and $\tilde{x}$ is the fraction of free volume required for rotor motion~\cite{gegiou_1968}. By combining these equations, F{\"o}rster-Hoffmann equation~\eqref{eq:FH_QY} can be retrieved with $x  =\tilde{x}/B$ and $C = \log \left( \varphi_0 A^{\tilde{x}/B} \right)$. As lifetime is directly proportional to the quantum yield, it follows:
\begin{equation}
\tau = \tau_0 e^{\tilde{x}/\nu}
\label{eq:free_vol_tau}
\end{equation}
\noindent in which $\tau_0$ is the fluorescence lifetime that the FMR would have if intramolecular rotation was fully free.

Based on this theory, FMRs were thus used as free volume probes in solid-state polymers to detect their glass transition~\cite{loutfy_1986,meyer_1990,hooker_1995,zhu_2007,jee_2009,hinze_2011,mirzahossein_2022,kim_2023e,jutze_2025}. In particular, local measurements in thin polymer films described an increase in free volume close to film surface~\cite{tanaka_2009,kawaguchi_2015} expected from theory. This qualitatively supports the theory of the free volume, but the relationship between microviscosity and free volume has never been directly and quantitatively tested.

The results reported in this study thus open an opportunity to bridge this gap, as the fluorescence lifetime $\tau \propto \varphi$ and density $\rho = 1/v_\text{m}$ were systematically measured. Qualitatively, it is known that the free volume is larger for small chains, due to the increasing contribution of extremities which are associated to larger fluctuations: this thus agrees with the observation of increasing lifetime $\tau$ with the proportion $W$ of heavy PEG in ternary mixtures.

\subsection{Direct use of free volume}

The main difficulty for quantitative test of the free volume theory lays in the estimation of the hard core volume $v_0$~\cite{haward_1970}. In some systems, such volume can be extrapolated from measured equations of state or by dilatometric techniques: for aqueous PEG solutions however, dilation factor in liquid state are tiny and measurements cannot be performed below $\SI{0}{\celsius}$, making these methods inapplicable. The best approach would be through numerical simulations, but this would go beyond the scope of this study. Alternatively, the hard core volume $v_0$ can be computed from group contribution approach: for a solution of two PEG with overall PEG fraction $w$ and proportion $W$ of one of the PEG (designated as 2), 
\begin{equation}
v_0 = (1-w) v_\text{w} + w \left[ (1-W) v_\text{PEG1} + W v_\text{PEG2} \right]
\end{equation}
\noindent in which $v_\text{w}$, $v_\text{PEG1}$ and $v_\text{PEG2}$ are respectively the specific hard core volumes of water and of the PEGs. For PEG molecules, group contributions methods can be used: in what follows, group parameters values by Sugden~\cite{sugden_1927b} were selected. It is to note that other values exist and can affect the result. More importantly, such methods are designed mostly for organic molecules and generally fail for associative molecules. They thus cannot be used for water and $v_\text{w}$ appears somehow as a free parameter.

\begin{figure*}[!htb]
    \includegraphics{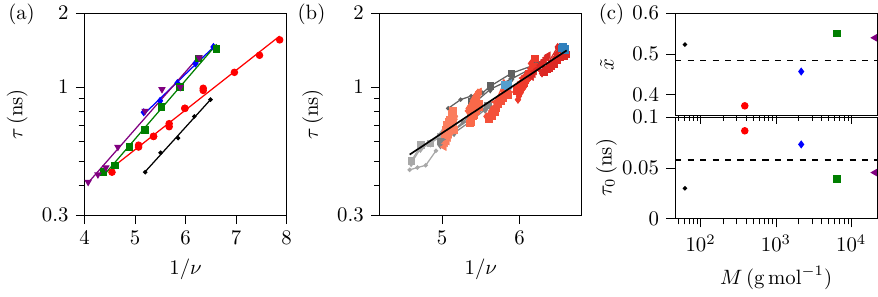}
    \caption{Direct test of free volume theory for binary and ternary mixtures. Continuous lines are fits along Eq.~\eqref{eq:free_vol_tau}. (a) Evolution of fluorescence lifetime $\tau$ with inverse free volume fraction $1/\nu$ for binary mixtures of water and PEG ({\tiny $ ^\bullet$} PEG-62; {\color{red} $\bullet$} PEG-400; {\tiny \color{blue} $^\blacklozenge$} PEG-2000; {\tiny \color{OliveGreen} $\blacksquare$} PEG-6000). (b) Evolution of fluorescence lifetime $\tau$ with inverse free volume fraction $1/\nu$ for ternary mixtures. Color corresponds to the lighter PEG (grey: PEG-62; red: PEG-400; blue: PEG-2000), color gradient corresponds to the average PEG fraction $w$ (darker color is associated with more concentrated solution), and symbols correspond to the heavier PEG ($\bullet$ PEG-400; {\tiny $^\blacklozenge$} PEG-2000; {\tiny $\blacksquare$} PEG-6000; {\tiny $\blacktriangledown$} PEG-20000). (c) Evolution of fit parameters $\tilde{x}$ and $\tau_0$ in Eq.~\eqref{eq:free_vol_tau} for binary mixtures ({\tiny $ ^\bullet$} PEG-62; {\color{red} $\bullet$} PEG-400; {\tiny \color{blue} $^\blacklozenge$} PEG-2000; {\tiny \color{OliveGreen} $\blacksquare$} PEG-6000). Value obtained \corr{from fit along Eq.~\eqref{eq:free_vol_tau}} for \corr{ternary} mixtures is represented by a dashed line.}
    \label{fig:freevolume_direct}
\end{figure*}

To determine an appropriate value of $v_\text{w}$, a first adjustment was performed using results on binary mixtures, depicted on Fig.~\ref{fig:freevolume_direct}(a). Eq.~\eqref{eq:free_vol_tau} predicts a linear relationship between the logarithmic lifetime $\ln \tau$ and the inverse free volume fraction $1/\nu$. A value of $v_\text{w} = \SI{0.776}{\centi\meter\cubed\per\gram}$ was obtained to obtain an optimal linear evolution. However, the coefficients $\tau_0$ and $\tilde{x}$, represented by points on Fig.~\ref{fig:freevolume_direct}(c), show noticeable variations with the considered PEG. This does not contradict application of free volume theory, yet a microscopic explanation of this evolution remains unclear.

Once a value of $v_\text{w}$ is selected from results on binary solutions, it is possible to test the validity of free volume theory on ternary mixtures. The obtained evolution of lifetime $\tau$ with inverse free volume fraction $1/\nu$ is displayed on Fig.~\ref{fig:freevolume_direct}(b), and corresponding fitting parameters are displayed as a dashed line on Fig.~\ref{fig:freevolume_direct}(c). All data of lifetime $\tau$ as a function of inverse free volume fraction $1/\nu$ collapse on a single curve, linear in semilogarithmic coordinates, with parameters close to these obtained with binary solution. This validates the direct applicability of free volume to the mixtures under study.

However, a closer inspection of data raises some possible doubts. First, it seems that data \corr{for the different average fractions $w$ regroup in a set of lines, with steeper slope and different intercepts. Instead of a single mastercurve, data would then be described as a series of powerlaw with different parameters $\tau_0$ and $\tilde{x}$, depending on average PEG fraction}. While this is not contradictory with the qualitative picture of free volume theory, it makes it less interesting for applications \corr{as systematic calibration of the evolution of lifetime $\tau$ with free volume $\nu$ would be required for any new mixture, even without changing significantly the chemical nature of its compounds}. Second, the results are highly sensitive to the chosen value of $v_\text{w}$ (and overall to the computation of the hard core volume $v_0$). For instance, changing the value of $v_\text{w}$ by a few percents can change drastically the shape of curves obtained on Fig.~\ref{fig:freevolume_direct}(a) and (b), breaking linearity of the evolution and enhancing this grouping effect of the mixtures of different molecular weights. In the absence of a more reliable method to estimate hard core volume $v_0$, this makes it difficult to unambiguously confirm the validity of the free volume theory.

\subsection{Local approach of free volume}

In the previous approach, free volume is considered as a homogeneous parameter, defined for the whole solution. However, typical gyration radius of PEG in water are of the order of a few nanometers~\cite{devanand_1991}, which can be compared to the characteristic molecular size of the FMR used in this study\corr{, while water molecules are notably smaller}. Consequently, in ternary mixtures, FMR molecules \corr{can be seen as interacting at a given time with only one of the two different types of PEG, in a bath of water molecules} This opens an alternative, local point of view on free volume theory. Instead of considering free volume as a homogeneous property of the solution, it is possible to consider that FMRs in solution can be in two different environments, corresponding to \corr{a homogeneous background of water molecules and in presence of} one or the other of the two PEGs, \corr{yet with an effective interaction which is influenced by the composition of the liquid at larger scale through the memory effect on microviscosity}. These environments correspond to the limit solutions $(1)$ and $(2)$, of local free volume $v_\text{m}^{(1/2)}$ leading to a lifetime $\tau^{(1/2)}$ as introduced above.  

As FLIM acquisition time (of the order of a few second) is significantly larger than molecular rearrangement time, and samples a large volume (for one pixel, the sampled volume is given by pixel linear field and lens depth of field, namely $\SI{2.5}{\micro\meter} \times \SI{2.5}{\micro\meter} \times \SI{7.5}{\micro\meter}$), the obtained lifetime is a volume average of these limit two environments (see SI for a more detailed explanation). Considering the small variations of density and proximity of density of the different solutions, the volume proportion of the two PEGs is equal to the mass ratio $W$ within a few percents: this thus justifies the observed ideal mixing rule, $\tau \simeq \tau^{(1)} (1-W_2) + \tau^{(2)} W_2$.

\begin{figure}[!htb]
    \includegraphics{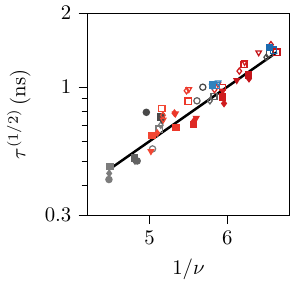}
    \caption{Test of local free volume theory for ternary mixtures: respective evolution of the lifetime $\tau^{(1/2)}$ defined in Eq.~\eqref{eq:ideal_ternary} as a function of inverse free volume fraction $1/\nu$ for the two limit solutions (filled/open symbols). Color corresponds to the lighter PEG (grey: PEG-62; red: PEG-400; blue: PEG-2000), color gradient corresponds to the average PEG fraction $w$ (darker color is associated with more concentrated solution), and symbols correspond to the heavier PEG ($\bullet$ PEG-400; {\tiny $^\blacklozenge$} PEG-2000; {\tiny $\blacksquare$} PEG-6000; {\tiny $\blacktriangledown$} PEG-20000). The continuous line correspond to a fit along Eq.~\eqref{eq:free_vol_tau}.}
    \label{fig:freevolume_extrapole}
\end{figure}

With this approach, free volume theory predicts that the lifetimes in limit solutions $\tau^{(1/2)}$ should evolve exponentially with the inverse free volume fractions $1/\nu$ computed from the specific volumes $v_\text{m}^{(1/2)}$. This hypothesis is tested on Fig.~\ref{fig:freevolume_extrapole}, and fit with Eq.~\eqref{eq:free_vol_tau} leads to $\tau_0 = \SI{6.5e-2}{\nano\second}$ and $\tilde{x} = 0.67$.

Experimental data seem to regroup satisfactorily on a single, linear curve with no clear grouping of data for solutions of different molecular weights, and without significant discrepancy for the two limit solutions $(1)$ and $(2)$. Moreover, modifying the value of $v_\text{w}$ within about $\SI{15}{\percent}$ affects values of $\tau_0$ and $\tilde{x}$ but does not modify qualitatively the look of these curves. Such results show that this alternative point of view is more robust than considering free volume as a global parameter.

\section{Conclusion}

While FMRs have emerged as promising local probes of microviscosity, their use in complex media, in particular biological, will remain limited to qualitative fluorescent sensing agents unless microviscosity is related to physico-chemical parameters. In this study, the evolution of the fluorescence lifetime of a carefully selected FMR has been investigated in both binary and ternary macromolecular mixtures of PEG and water.

In binary mixtures, the results are overall consistent with those  reported by Bittermann et al.~\cite{bittermann_2021a}: while F{\"o}rster-Hoffmann equation describes satisfactorily the evolution of lifetime with viscosity, the fitting parameters depend on the nature of the solution, showing that microviscosity is not solely dictated by viscosity. However, mass fraction of PEG is also not sufficient to predict lifetime.

In ternary mixtures, in the range of compositions investigated, ideal mixing rules were observed for the logarithm of viscosity $\eta$, for specific volumes $v_\text{m}$, and for fluorescent lifetimes $\tau$, which leads to the definition of two limit binary solutions, whose properties are obtained by extrapolating measurements for ternary mixtures. These measurements finally allowed to test quantitatively the free volume theory. A first, direct approach is consistent with the obtained results, yet it requires to finely tune the value of the hard core specific volume $v_0$, an empirical parameter involved in the definition of free volume. Alternatively, by considering that the two limit solutions defined previously correspond to two possible micro-environments around the FMR, the linear mixing rule observed for lifetime can be justified, and free volume theory is validated for these limit solutions. Such approach is more robust to the value of $v_0$ and opens a new way of thinking about free volume and microviscosity in complex environments.

Yet, work remains to be done to fully understand the parameters ruling microviscosity. \corr{The experiments presented here investigated complex microenvironments, for which not only the zero-frequency viscosity is modified, but also the frequency-dependent memory effect of microfriction that has been proved to influence microscopic transition rates.~\cite{northrup_1980,grote_1980,dalton_2024}} While free volume theory appears \corr{as a promising phenomenological approach}, it should still be tested in solutions that deviate from ideal behavior, and of different chemical natures. This effort would however be necessary to fully exploit the potential of FMRs in local characterization of complex media.

\acknowledgements{The authors thank Oualid Mekkaoui and Arthur Roudaut for their help in the synthesis of the FMR. They also acknowledge Thomas Salez, Philippe Marchal and Chloé Grazon for discussions on different aspects of this work. Christophe Velours and Frédérique Ham-Pichavant of the polymer characterization platform PolyCar at LCPO are thanked for analysis of samples molecular weight distribution. MC was funded by the China Scholarship Council. \corr{This project has received financial support from the CNRS through the MITI interdisciplinary programs through its exploratory research program.}}

\newpage

\appendix

\corr{

\section{Principle of frequency-domain FLIM}

The frequency-domain FLIM technique used in this article measures fluorescence lifetime by rather indirect method: the main features of signal analysis are described in this section, and details can be found for instance in Ref.~\cite{lakowicz}

The fluorescent probe is excited by a modulated signal of intensity $I_\text{exc}$:
\begin{equation}
    I_\text{exc} (t) = I_0 \left[1+M_\text{exc} \cos(\omega t)\right],
\end{equation}
\noindent in which the modulation pulsation $\omega$ and the modulation depth $M_\text{exc}$ are controlled by the light source. In the linear regime, the resulting fluorescence signal $I_\text{em}$ is modulated at the same frequency and is given by:
\begin{equation}
    I_\text{em} (t) = I_0 \left[1+M_\text{em} \cos(\omega t + \varphi)\right].
\end{equation}
For a mono-exponential relaxation, the modulation depth and phase delay of the emitted signal are related to lifetime $\tau$ through:
\begin{equation}
    M_\text{em} = \frac{M_\text{exc}}{\sqrt{1+(\omega \tau)^2}} \quad \text{and}\quad \varphi = - \mathrm{Arctan}(\omega \tau).
\end{equation}
\noindent As the absolute origin of time is not known, determination of $M_\text{em}$ and $\varphi$ are not direct. They are obtained with a lock-in amplifier: the collected fluorescence signal is multiplied by a controlled reference signal $D(t) = D_0 \left[1+M_\text{ref} \cos (\omega t + \varphi_\text{D}) \right]$ and the result is low-pass filtered. The resulting signal is stationary and proportional to:
\begin{equation}
    S \propto S_0 \left[ 1+ \frac{M_\text{em} M_\text{ref}}{2} \cos (\varphi - \varphi_\text{D} ) \right].
    \label{eq:FLIM_lock_in}
\end{equation}
\noindent By measuring the signal for different values of the imposed reference phase $\varphi_\text{D}$, and adjusting the result by Eq.~\eqref{eq:FLIM_lock_in}, $M_\text{em}$ and $\varphi$ can be obtained.

Lifetime can thus be retrieved from both the modulation depth and the phase delay: in practice, the latter is less sensitive to noise and was systematically used in the article. Finally, this description hides the fact that the intensities are defined at the probe molecule, and amplitude and phase can be modified by electronic acquisition and optical path: reference measurement performed in similar conditions allows to factor out these effects.}

\section{Description of NMR spectra}

$^1$H NMR (300 MHz, CDCl$_3$) $\delta$ 7.92 – 7.85 (AA'XX', $J_\mathrm{AA'/XX'} = \SI{1.98}{\hertz}$, $J_\mathrm{AX} = \SI{7.81}{\hertz}$, $J_\mathrm{AX'} = \SI{0.34}{\hertz}$, 2H), 7.82 (d, $J = \SI{2.1}{\hertz}$, 1H), 7.53 – 7.43 (AA'XX', $J_\mathrm{AA'/XX'} = \SI{1.98}{\hertz}$, $J_\mathrm{AX} = \SI{7.81}{\hertz}$, $J_\mathrm{AX'} = \SI{0.34}{\hertz}$, 2H), 7.37 – 7.28 (m, 4H), 6.99 (d, $J = \SI{4.4}{\hertz}$, 1H), 6.86 (d, $J = \SI{4.1}{\hertz}$, 1H), 6.72 – 6.65 (m, 1H), 6.50 (dd, $J = 4.3$, \SI{1.9}{\hertz}, 1H), 2.48 (s, 3H), 2.43 (s, 3H).

\section{List of solutions}

\begin{table}[!htb]
\centering
\begin{tabular}{|c|c|c|c|c|c|}
\hline
\backslashbox{$M_1$}{$M_2$} & 62 & 400 & 2000 & 6000 & 20000 \\ \hline
62 & 0.6 $\stackrel{0.1}{\rightarrow}$ 1 $^\star$ &  0.4 $\stackrel{0.1}{\rightarrow}$ 0.6 $^\star$ &  0.4 $\stackrel{0.1}{\rightarrow}$ 0.6 $^\star$ &  0.4 $\stackrel{0.1}{\rightarrow}$ 0.6 $^\star$ & $\emptyset$ \\ \hline
400 & - & 0.4 $\stackrel{0.05}{\rightarrow}$ 0.6 & 0.4 $\stackrel{0.05}{\rightarrow}$ 0.6 & 0.4 $\stackrel{0.05}{\rightarrow}$ 0.6 & 0.4 $\stackrel{0.05}{\rightarrow}$ 0.6 \\ \hline
2000 & - & - & 0.4 $\stackrel{0.05}{\rightarrow}$ 0.6 & 0.5 $\stackrel{0.1}{\rightarrow}$ 0.6 $^\star$ & 0.5 $^\star$ \\ \hline
6000 & - & - & - & 0.5 $\stackrel{0.05}{\rightarrow}$ 0.6 & $\emptyset$ \\ \hline
20000 & - & - & - & - & 0.14 $\rightarrow$ 0.55 \\
\hline
\end{tabular} 
\caption{List of investigated solutions of water, a light PEG of molecular weight $M_1$ and a heavier PEG of molecular weight $M_2$. Table contains the range of investigated overall PEG fraction $w$ (with step indicated as superscript). Diagonal cells correspond to binary mixtures, - indicate symmetric terms, and $\emptyset$ indicate solutions that were not considered. For all ternary mixtures, different solutions of varying proportion $W$ of heavy PEG were prepared, with $W$ between $0.2$ and $0.8$ by steps of $0.2$ (cells with a $^\star$) or between $0.1$ and $0.9$ by steps of $0.1$.} 
\label{tab:table1}
\end{table}

\section{Fit parameters for binary solutions}

Properties of binary mixtures of PEG of molecular weight $M$ and PEG mass fraction $w$, represented on Fig.3--4 in main text, were fitted along:
\begin{equation}
\left\{ \begin{array}{l}
v_\text{m} (M,w) = \alpha_v(M)w^2 + \beta_v(M) w + \gamma_v(M) \\
\ln \eta(w,M) = (1-w) \ln \eta_0(M) + w \ln \eta_1(M) \\ 
\ln \tau(w,M) = (1-w) \ln \tau_0(M) + w \ln \tau_1(M)
\end{array} \right. .
\label{eq:fit_binary}
\end{equation}
\noindent Values of the different parameters are gathered in Table~\ref{tab:table2}.

\begin{table}[!htb]
\centering
\begin{equation*}
\begin{array}{|c||c|c|c||c|c||c|c|}
\hline
M & \alpha_v & \beta_v & \gamma_v & \ln\eta_0 & \ln \eta_1 & \ln\tau_0 & \ln\tau_1 \\ \hline
62 & 0.04 & -0.13 & 1.00 & -0.41 & 2.20 & -1.82 & -0.12 \\ \hline
400 & 0.13 & -0.27 & 1.02 & 0.32 & 4.10 & -1.41 & 0.75 \\ \hline
2000 & 0.13 & -0.28 & 1.02 & 0.24 & 7.48 & -1.52 & 1.63 \\ \hline
6000 & 0.08 & -0.22 & 1.01 & 0.94 & 9.07 & -1.78 & 1.80  \\ \hline
20000 & 0.05 & -0.20 & 1.01 & 1.65 & 13.65 & -1.46 & 1.60 \\ \hline
\end{array}
\end{equation*}
\caption{List of fitting parameters for binary mixtures, along Eq.~\eqref{eq:fit_binary}. Molecular weight $M$ are expressed in $\SI{}{\gram\per\mole}$; parameters $\alpha_v$, $\beta_v$ and $\gamma_v$ are expressed in $\SI{}{\gram\per\centi\meter\cubed}$; parameters $\eta_0$ and $\eta_1$ are expressed in $\SI{}{\milli\pascal\second}$; parameters $\tau_0$ and $\tau_1$ are expressed in $\SI{}{\nano\second}$. Note that in Eq.~\eqref{eq:fit_binary}, $w$ is expressed with no unit (in particular it should not be converted in $\SI{}{\percent}$).}
\label{tab:table2}
\end{table}

\section{Effect of temperature on fluorescence lifetime}

In order to investigate the effect of temperature on the lifetime of the FMR, several aqueous binary mixtures of PEG-2000 with mass fraction in range $w=\SIrange{0.4}{0.6}{}$ were prepared. Their viscosity were measured as described in main text, using Peltier element of the rheometer to impose the temperature. Lifetimes were measured using a temperature-controlled sample holder mounted in an Edinburgh FLS1000 spectrofluorometer equipped with an HS-PMT-870 detector with enhanced temporal response. The minimal fluorescence lifetime that can be accessed using this setup was $\SI{25}{\pico\second}$.

\begin{figure*}[!htb]
    \includegraphics{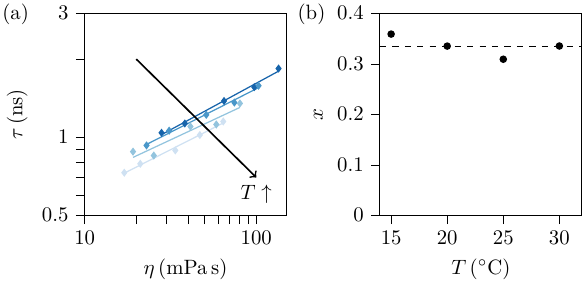}
    \caption{Effect of temperature on response of FMR. (a) Evolution of lifetime $\tau$ with viscosity $\eta$ in binary solutions of PEG-2000 of different mass fraction $w$, and at different temperatures $T$. Continuous lines correspond to fit with F{\"o}rster-Hoffmann equation. (b) Evolution of F{\"o}rster-Hoffmann exponent $x$ with temperature.}
    \label{fig:SI_temperature}
\end{figure*}

The evolution of lifetime $\tau$ with viscosity $\eta$ at different temperatures in range $T=\SIrange{15}{30}{\celsius}$ is displayed on Fig.~\ref{fig:SI_temperature}(a). On the whole temperature range, F{\"o}rster-Hoffmann equation is valid, with an exponent $x$ that is independent of temperature as can be seen on Fig.~\ref{fig:SI_temperature}(b). Yet, lifetime shows a clear dependency with temperature (through the coefficient $C_\tau$): in the experiments presented in main text, temperature is controlled within a few $\SI{}{\celsius}$ so possible fluctuations have no significant impact, but it is interesting to note that temperature effect are not a priori negligible when working with FMR.

\section{Data for all ternary mixtures involving PEG-400}

\begin{figure*}[!htb]
    \includegraphics{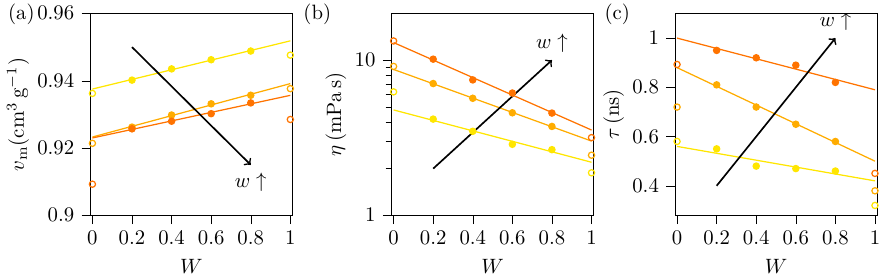}
    \caption{Properties of ternary mixture of water, PEG-62 and PEG-400 as a function of the proportion of PEG-62, $W$. The color gradient represents the evolution of the total PEG mass fraction $w$, varying between $w=0.4$ and $0.6$ by step of $0.1$. The open symbols on axis $W=0$ and $W=1$ represent data obtained in binary PEG/water mixtures, displayed on Fig.3--4 of main text. Continuous lines are linear fit along Eq.~\eqref{eq:ideal_ternary_SI}. \corr{(a) Evolution of specific volume $v_\text{m}$. (b) Evolution of viscosity $\eta$ in semilogarithmic coordinates.} (c) Evolution of lifetime $\tau$.}
    \label{fig:SI_example_ternary_400_62}
\end{figure*}

\begin{figure*}[!htb]
    \includegraphics{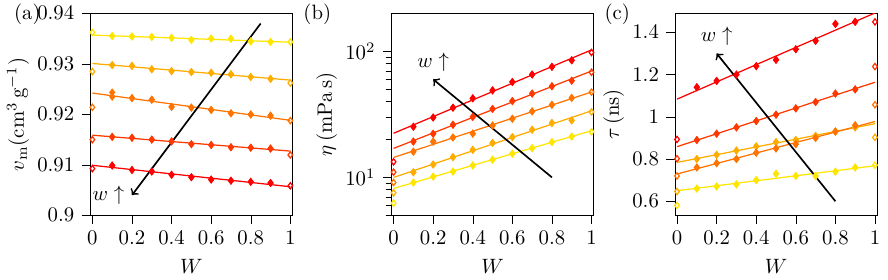}
    \caption{Properties of ternary mixture of water, PEG-400 and PEG-2000 as a function of the proportion of PEG-2000, $W$. The color gradient represents the evolution of the total PEG mass fraction $w$, varying between $w=0.4$ and $0.6$ by step of $0.05$. The open symbols on axis $W=0$ and $W=1$ represent data obtained in binary PEG/water mixtures, displayed on Fig.3--4 of main text. Continuous lines are linear fit along Eq.~\eqref{eq:ideal_ternary_SI}. \corr{(a) Evolution of specific volume $v_\text{m}$. (b) Evolution of viscosity $\eta$ in semilogarithmic coordinates.} (c) Evolution of lifetime $\tau$. (Same figure as Fig.~5 in main text)}
    \label{fig:SI_example_ternary_400_2000}
\end{figure*}

\begin{figure*}[!htb]
    \includegraphics{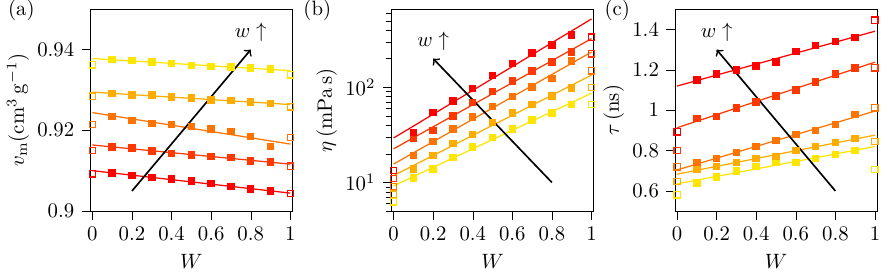}
    \caption{Properties of ternary mixture of water, PEG-400 and PEG-6000 as a function of the proportion of PEG-6000, $W$. The color gradient represents the evolution of the total PEG mass fraction $w$, varying between $w=0.4$ and $0.6$ by step of $0.05$. The open symbols on axis $W=0$ and $W=1$ represent data obtained in binary PEG/water mixtures, displayed on Fig.3--4 of main text. Continuous lines are linear fit along Eq.~\eqref{eq:ideal_ternary_SI}. \corr{(a) Evolution of specific volume $v_\text{m}$. (b) Evolution of viscosity $\eta$ in semilogarithmic coordinates.} (c) Evolution of lifetime $\tau$.}
    \label{fig:SI_example_ternary_400_6000}
\end{figure*}

\begin{figure*}[!htb]
    \includegraphics{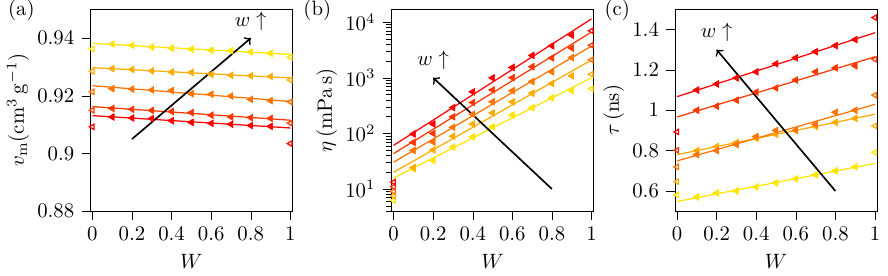}
    \caption{Properties of ternary mixture of water, PEG-400 and PEG-20000 as a function of the proportion of PEG-20000, $W$. The color gradient represents the evolution of the total PEG mass fraction $w$, varying between $w=0.4$ and $0.6$ by step of $0.05$. The open symbols on axis $W=0$ and $W=1$ represent data obtained in binary PEG/water mixtures, displayed on Fig.3--4 of main text. Continuous lines are linear fit along Eq.~\eqref{eq:ideal_ternary_SI}. \corr{(a) Evolution of specific volume $v_\text{m}$. (b) Evolution of viscosity $\eta$ in semilogarithmic coordinates.} (c) Evolution of lifetime $\tau$.}
    \label{fig:SI_example_ternary_400_20000}
\end{figure*}

The mixture of PEG-400 (designated as PEG 1) with other PEGs (designated as PEG 2) was the most systematically investigated in this study: Fig.~\ref{fig:SI_example_ternary_400_62},~\ref{fig:SI_example_ternary_400_2000} (similar to Fig.~5 in main text), ~\ref{fig:SI_example_ternary_400_6000} and~\ref{fig:SI_example_ternary_400_20000} display results of evolution of specific volume, viscosity and lifetime for the respective mixtures with PEG-62, PEG-6000 and PEG-20000, as a function of the mass proportion $W$ of PEG 2. Open symbols at $W=0$ and $W=1$ are measurements performed for corresponding binary mixtures displayed on Fig.3--4 of main text, when available. As discussed in main text, lines correspond to fits with ideal mixing rules:
\begin{widetext}
\begin{equation}
\left\{\begin{array}{lll}
v_\text{m} (M_1,M_2,w,W_2) &= v^{(1)}_\text{m} (M_1,M_2,w) (1-W_2) &+ v^{(2)}_\text{m} (M_1,M_2,w) W_2 \\
\ln \eta (M_1,M_2,w,W) &= \ln \eta^{(1)} (M_1,M_2,w) (1-W_2) &+ \ln \eta^{(2)} (M_1,M_2,w) W_2 \\
\tau (M_1,M_2,w,W_2) &= \tau^{(1)} (M_1,M_2,w) (1-W_2) &+ \tau^{(2)} (M_1,M_2,w) W_2
\end{array} \right. .
\label{eq:ideal_ternary_SI}
\end{equation}
\end{widetext}
\noindent Qualitatively, results are similar to these for the PEG-400/PEG-2000 mixture discussed in main text.

\newpage

\section{Properties of limit solutions of ternary mixtures}

\begin{figure*}[!htb]
    \includegraphics{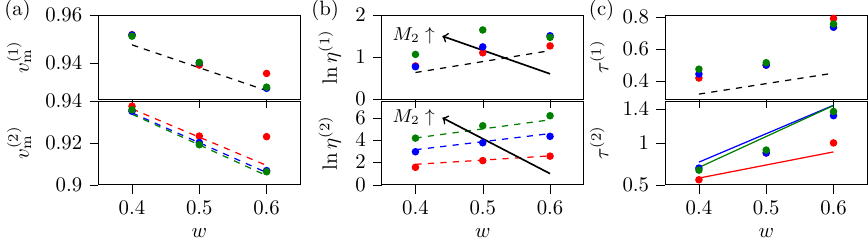}
    \caption{Evolution of fitting parameters in Eq.~\eqref{eq:ideal_ternary_SI}, corresponding to limit binary solutions $(1)$ (top row) and $(2)$ (bottom row), as a function of total PEG mass fraction $w$ and for mixtures of PEG-62 $(1)$ with another PEG $(2)$ ({\color{red} $\bullet$} PEG-400, {\color{blue} $\bullet$} PEG-2000, {\color{OliveGreen} $\bullet$} PEG-6000).  Dashed line corresponds to properties of binary solutions of water and PEG. (a) Specific volumes $v_\text{m}$, in $\SI{}{\centi\meter\cubed\per\gram}$. (b) Logarithm of viscosity $\ln \eta$ in $\SI{}{\milli\pascal\second}$. (c) Lifetime $\tau$, in $\SI{}{\nano\second}$.}
    \label{fig:SI_parameters_ternary_62_tot}
\end{figure*}

Values of the different fitting parameters in Eq.~\eqref{eq:ideal_ternary_SI} are given in following tables~\ref{tab:table_ternary_PEG62},~\ref{tab:table_ternary_PEG400},~\ref{tab:table_ternary_PEG2000},~\ref{tab:table_ternary_PEG6000} and ~\ref{tab:table_ternary_PEG20000}, and their evolution with the overall PEG mass fraction $w$ are displayed on Fig.~\ref{fig:SI_parameters_ternary_62_tot},~\ref{fig:SI_parameters_ternary_2000_tot},~\ref{fig:SI_parameters_ternary_6000_tot} and ~\ref{fig:SI_parameters_ternary_20000_tot}. Only values of parameters for limit solution $(1)$ are given, these for limit solution $(2)$ can be deduced by:
\begin{equation}
X^{(1)} (M_1,M_2,w) = X^{(2)} (M_2,M_1,w)
\end{equation}

\begin{table}[!htb]
\centering
\begin{tabular}{|cc|c|c|c|}
\hline
\multicolumn{2}{|c|}{\backslashbox{$M_2$}{$w$}} & 0.4 & 0.5 & 0.6   \\ \hline
\multicolumn{1}{|c|}{\multirow{3}{*}{400}} & $v_\text{m}^{(1)}$ & 0.952 & 0.939 & 0.936    \\ \cline{2-5} 
\multicolumn{1}{|c|}{}                  & $\ln \eta^{(1)}$ & 0.79 & 1.10 & 1.27 \\ \cline{2-5} 
\multicolumn{1}{|c|}{}                  & $\tau^{(1)}$ & 0.42 & 0.5 & 0.79    \\ \hline
\multicolumn{1}{|c|}{\multirow{3}{*}{2000}} & $v_\text{m}^{(1)}$ & 0.952 & 0.940 & 0.929  \\ \cline{2-5} 
\multicolumn{1}{|c|}{}                  & $\ln \eta^{(1)}$ & 0.77 & 1.24 & 1.51 \\ \cline{2-5} 
\multicolumn{1}{|c|}{}                  & $\tau^{(1)}$ & 0.445 & 0.50 & 0.735    \\ \hline
\multicolumn{1}{|c|}{\multirow{3}{*}{6000}} & $v_\text{m}^{(1)}$ & 0.951 & 0.940 & 0.930 \\ \cline{2-5} 
\multicolumn{1}{|c|}{}                  & $\ln \eta^{(1)}$ & 1.06 & 1.65 & 1.47 \\ \cline{2-5} 
\multicolumn{1}{|c|}{}                  & $\tau^{(1)}$ & 0.475 & 0.515 & 0.755   \\ \hline
\end{tabular}
\caption{List of fitting parameters for ternary mixtures for solutions of PEG-62 $(1)$ with another PEG $(2)$, along Eq.~\eqref{eq:ideal_ternary_SI}. Specific volumes are expressed in $\SI{}{\gram\per\centi\meter\cubed}$, viscosities are expressed in $\SI{}{\milli\pascal\second}$ and lifetimes are expressed in $\SI{}{\nano\second}$. Note that in Eq.~\eqref{eq:ideal_ternary_SI}, $w$ is expressed with no unit (in particular it should not be converted in $\SI{}{\percent}$).}
\label{tab:table_ternary_PEG62}
\end{table}

\begin{table}[!htb]
\centering
\begin{tabular}{|cc|c|c|c|c|c|}
\hline
\multicolumn{2}{|c|}{\backslashbox{$M_2$}{$w$}} & 0.4 & 0.45 & 0.5 & 0.55 & 0.6  \\ \hline
\multicolumn{1}{|c|}{\multirow{3}{*}{62}} & $v_\text{m}^{(1)}$ & 0.938 & - & 0.923 & - & 0.923 \\ \cline{2-7} 
\multicolumn{1}{|c|}{}                  & $\ln \eta^{(1)}$ & 1.56 & - & 2.16 & - & 2.56 \\ \cline{2-7} 
\multicolumn{1}{|c|}{}                  & $\tau^{(1)}$ & 0.56 & - & 0.88 & - & 1 \\ \hline
\multicolumn{1}{|c|}{\multirow{3}{*}{2000}} & $v_\text{m}^{(1)}$ & 0.936 & 0.930 & 0.924 & 0.916 & 0.910 \\ \cline{2-7} 
\multicolumn{1}{|c|}{}                  & $\ln \eta^{(1)}$ & 2.10 & 2.30 & 2.67 & 2.83 & 3.11 \\ \cline{2-7} 
\multicolumn{1}{|c|}{}                  & $\tau^{(1)}$ & 0.65 & 0.78 & 0.73 & 0.86 & 1.08 \\ \hline
\multicolumn{1}{|c|}{\multirow{3}{*}{6000}} & $v_\text{m}^{(1)}$ & 0.938 & 0.929 & 0.924 & 0.916 & 0.910 \\ \cline{2-7} 
\multicolumn{1}{|c|}{}                  & $\ln \eta^{(1)}$ & 2.22 & 2.47 & 2.75 & 3.12 & 3.38 \\ \cline{2-7} 
\multicolumn{1}{|c|}{}                  & $\tau^{(1)}$ & 0.64 & 0.68 & 0.70 & 0.94 & 1.12 \\ \hline
\multicolumn{1}{|c|}{\multirow{3}{*}{20000}} & $v_\text{m}^{(1)}$ & 0.938 & 0.930 & 0.924 & 0.916 & 0.913 \\ \cline{2-7} 
\multicolumn{1}{|c|}{}                  & $\ln \eta^{(1)}$ & 2.76 & 3.00 & 3.41 & 3.77 & 4.11 \\ \cline{2-7} 
\multicolumn{1}{|c|}{}                  & $\tau^{(1)}$ & 0.55 & 0.78 & 0.75 & 0.97 & 1.07 \\ \hline
\end{tabular}
\caption{List of fitting parameters for ternary mixtures for solutions of PEG-400 $(1)$ with another PEG $(2)$, along Eq.~\eqref{eq:ideal_ternary_SI}. Specific volumes are expressed in $\SI{}{\gram\per\centi\meter\cubed}$, viscosities are expressed in $\SI{}{\milli\pascal\second}$ and lifetimes are expressed in $\SI{}{\nano\second}$. Note that in Eq.~\eqref{eq:ideal_ternary_SI}, $w$ is expressed with no unit (in particular it should not be converted in $\SI{}{\percent}$).}
\label{tab:table_ternary_PEG400}
\end{table}

\begin{figure*}[!htb]
    \includegraphics{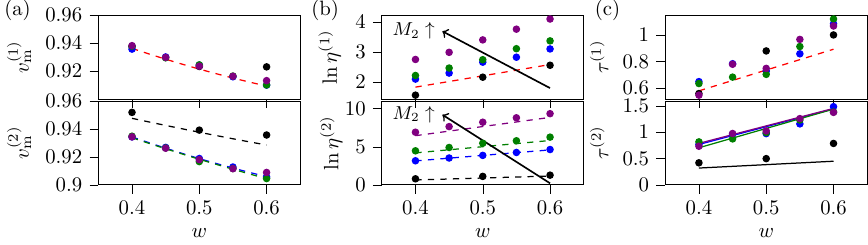}
    \caption{Evolution of fitting parameters in Eq.~\eqref{eq:ideal_ternary_SI}, corresponding to limit binary solutions $(1)$ (top row) and $(2)$ (bottom row), as a function of total PEG mass fraction $w$ and for mixtures of PEG-400 $(1)$ with another PEG $(2)$ ({$\bullet$} PEG-62, {\color{blue} $\bullet$} PEG-2000, {\color{OliveGreen} $\bullet$} PEG-6000, {\color{violet} $\bullet$} PEG-20000).  Dashed line corresponds to properties of binary solutions of water and PEG. (a) Specific volumes $v_\text{m}$, in $\SI{}{\centi\meter\cubed\per\gram}$. (b) Logarithm of viscosity $\ln \eta$ in $\SI{}{\milli\pascal\second}$. (c) Lifetime $\tau$, in $\SI{}{\nano\second}$. (Same figure as Fig.~6 in main text)}
    \label{fig:SI_parameters_ternary_400_tot}
\end{figure*}

\begin{table}[!htb]
\centering
\begin{tabular}{|cc|c|c|c|c|c|}
\hline
\multicolumn{2}{|c|}{\backslashbox{$M_2$}{$w$}} & 0.4 & 0.45 & 0.5 & 0.55 & 0.6  \\ \hline
\multicolumn{1}{|c|}{\multirow{3}{*}{62}} & $v_\text{m}^{(1)}$ & 0.935 & - & 0.920 & - & 0.907 \\ \cline{2-7} 
\multicolumn{1}{|c|}{}                  & $\ln \eta^{(1)}$ & 2.96 & - & 3.76 & - & 4.34 \\ \cline{2-7} 
\multicolumn{1}{|c|}{}                  & $\tau^{(1)}$ & 0.7 & - & 0.88 & - & 1.33 \\ \hline
\multicolumn{1}{|c|}{\multirow{3}{*}{400}} & $v_\text{m}^{(1)}$ & 0.934 & 0.927 & 0.919 & 0.913 & 0.906 \\ \cline{2-7} 
\multicolumn{1}{|c|}{}                  & $\ln \eta^{(1)}$ & 3.16 & 3.52 & 3.86 & 4.25 & 4.63 \\ \cline{2-7} 
\multicolumn{1}{|c|}{}                  & $\tau^{(1)}$ & 0.77 & 0.97 & 0.98 & 1.16 & 1.49 \\ \hline
\multicolumn{1}{|c|}{\multirow{3}{*}{6000}} & $v_\text{m}^{(1)}$ & - & - & 0.919 & - & 0.906 \\ \cline{2-7} 
\multicolumn{1}{|c|}{}                  & $\ln \eta^{(1)}$ & - & - & 3.92 & - & 4.67 \\ \cline{2-7} 
\multicolumn{1}{|c|}{}                  & $\tau^{(1)}$ & - & - & 1.02 & - & 1.46 \\ \hline
\multicolumn{1}{|c|}{\multirow{3}{*}{20000}} & $v_\text{m}^{(1)}$ & - & - & 0.920 & - & - \\ \cline{2-7} 
\multicolumn{1}{|c|}{}                  & $\ln \eta^{(1)}$ & - & - & 4.64 & - & - \\ \cline{2-7} 
\multicolumn{1}{|c|}{}                  & $\tau^{(1)}$ & - & - & 1.01 & - & - \\ \hline
\end{tabular}
\caption{List of fitting parameters for ternary mixtures for solutions of PEG-2000 $(1)$ with another PEG $(2)$, along Eq.~\eqref{eq:ideal_ternary_SI}. Specific volumes are expressed in $\SI{}{\gram\per\centi\meter\cubed}$, viscosities are expressed in $\SI{}{\milli\pascal\second}$ and lifetimes are expressed in $\SI{}{\nano\second}$. Note that in Eq.~\eqref{eq:ideal_ternary_SI}, $w$ is expressed with no unit (in particular it should not be converted in $\SI{}{\percent}$).}
\label{tab:table_ternary_PEG2000}
\end{table}

\begin{figure*}[!htb]
    \includegraphics{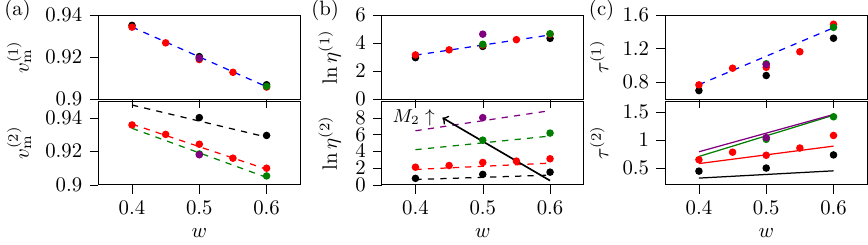}
    \caption{Evolution of fitting parameters in Eq.~\eqref{eq:ideal_ternary_SI}, corresponding to limit binary solutions $(1)$ (top row) and $(2)$ (bottom row), as a function of total PEG mass fraction $w$ and for mixtures of PEG-2000 $(1)$ with another PEG $(2)$  ($\bullet$ PEG-62, {\color{red} $\bullet$} PEG-400, {\color{OliveGreen} $\bullet$} PEG-6000, {\color{violet} $\bullet$} PEG-20000).  Dashed line corresponds to properties of binary solutions of water and PEG. (a) Specific volumes $v_\text{m}$, in $\SI{}{\centi\meter\cubed\per\gram}$. (b) Logarithm of viscosity $\ln \eta$ in $\SI{}{\milli\pascal\second}$. (c) Lifetime $\tau$, in $\SI{}{\nano\second}$.}
    \label{fig:SI_parameters_ternary_2000_tot}
\end{figure*}

\begin{table}[!htb]
\centering
\begin{tabular}{|cc|c|c|c|c|c|}
\hline
\multicolumn{2}{|c|}{\backslashbox{$M_2$}{$w$}} & 0.4 & 0.45 & 0.5 & 0.55 & 0.6  \\ \hline
\multicolumn{1}{|c|}{\multirow{3}{*}{62}} & $v_\text{m}^{(1)}$ & 0.936 & - & 0.919 & - & 0.906 \\ \cline{2-7} 
\multicolumn{1}{|c|}{}                  & $\ln \eta^{(1)}$ & 4.18 & - & 5.29 & - & 6.18 \\ \cline{2-7} 
\multicolumn{1}{|c|}{}                  & $\tau^{(1)}$ & 0.68 & - & 0.92 & - & 1.38 \\ \hline
\multicolumn{1}{|c|}{\multirow{3}{*}{400}} & $v_\text{m}^{(1)}$ & 0.935 & 0.926 & 0.917 & 0.912 & 0.904 \\ \cline{2-7} 
\multicolumn{1}{|c|}{}                  & $\ln \eta^{(1)}$ & 4.46 & 4.90 & 5.46 & 5.78 & 6.26 \\ \cline{2-7} 
\multicolumn{1}{|c|}{}                  & $\tau^{(1)}$ & 0.82 & 0.88 & 1.00 & 1.24 & 1.39 \\ \hline
\multicolumn{1}{|c|}{\multirow{3}{*}{2000}} & $v_\text{m}^{(1)}$ & - & - & 0.918 & - & 0.905 \\ \cline{2-7} 
\multicolumn{1}{|c|}{}                  & $\ln \eta^{(1)}$ & - & - & 5.33 & - & 6.18 \\ \cline{2-7} 
\multicolumn{1}{|c|}{}                  & $\tau^{(1)}$ & - & - & 1.02 & - & 1.42 \\ \hline
\end{tabular}
\caption{List of fitting parameters for ternary mixtures for solutions of PEG-6000 $(1)$ with another PEG $(2)$, along Eq.~\eqref{eq:ideal_ternary_SI}. Specific volumes are expressed in $\SI{}{\gram\per\centi\meter\cubed}$, viscosities are expressed in $\SI{}{\milli\pascal\second}$ and lifetimes are expressed in $\SI{}{\nano\second}$. Note that in Eq.~\eqref{eq:ideal_ternary_SI}, $w$ is expressed with no unit (in particular it should not be converted in $\SI{}{\percent}$).}
\label{tab:table_ternary_PEG6000}
\end{table}

\begin{figure*}[!htb]
    \includegraphics{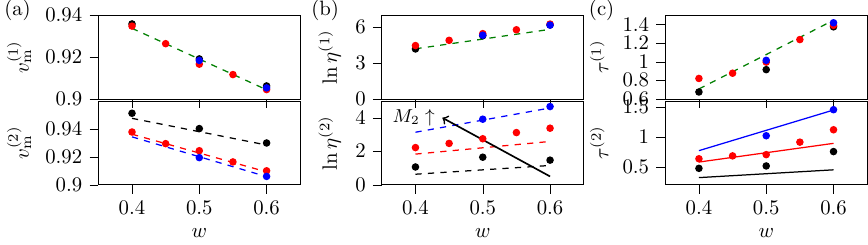}
    \caption{Evolution of fitting parameters in Eq.~\eqref{eq:ideal_ternary_SI}, corresponding to limit binary solutions $(1)$ (top row) and $(2)$ (bottom row), as a function of total PEG mass fraction $w$ and for mixtures of PEG-6000 $(1)$ with another PEG $(2)$ ($\bullet$ PEG-62, {\color{red} $\bullet$} PEG-400, {\color{blue} $\bullet$} PEG-2000).  Dashed line corresponds to properties of binary solutions of water and PEG. (a) Specific volumes $v_\text{m}$, in $\SI{}{\centi\meter\cubed\per\gram}$. (b) Logarithm of viscosity $\ln \eta$ in $\SI{}{\milli\pascal\second}$. (c) Lifetime $\tau$, in $\SI{}{\nano\second}$.}
    \label{fig:SI_parameters_ternary_6000_tot}
\end{figure*}

\begin{table}[!htb]
\centering
\begin{tabular}{|cc|c|c|c|c|c|}
\hline
\multicolumn{2}{|c|}{\backslashbox{$M_2$}{$w$}} & 0.4 & 0.45 & 0.5 & 0.55 & 0.6  \\ \hline
\multicolumn{1}{|c|}{\multirow{3}{*}{400}} & $v_\text{m}^{(1)}$ & 0.934 & 0.926 & 0.918 & 0.912 & 0.909 \\ \cline{2-7} 
\multicolumn{1}{|c|}{}                  & $\ln \eta^{(1)}$ & 6.91 & 7.64 & 8.22 & 8.80 & 9.34 \\ \cline{2-7} 
\multicolumn{1}{|c|}{}                  & $\tau^{(1)}$ & 0.74 & 0.98 & 1.03 & 1.26 & 1.38 \\ \hline
\multicolumn{1}{|c|}{\multirow{3}{*}{2000}} & $v_\text{m}^{(1)}$ & - & - & 0.918 & - & - \\ \cline{2-7} 
\multicolumn{1}{|c|}{}                  & $\ln \eta^{(1)}$ & - & - & 8.02 & - & - \\ \cline{2-7} 
\multicolumn{1}{|c|}{}                  & $\tau^{(1)}$ & - & - & 1.05 & - & - \\ \hline
\end{tabular}
\caption{List of fitting parameters for ternary mixtures for solutions of PEG-20000 $(1)$ with another PEG $(2)$, along Eq.~\eqref{eq:ideal_ternary_SI}. Specific volumes are expressed in $\SI{}{\gram\per\centi\meter\cubed}$, viscosities are expressed in $\SI{}{\milli\pascal\second}$ and lifetimes are expressed in $\SI{}{\nano\second}$. Note that in Eq.~\eqref{eq:ideal_ternary_SI}, $w$ is expressed with no unit (in particular it should not be converted in $\SI{}{\percent}$).}
\label{tab:table_ternary_PEG20000}
\end{table}

\begin{figure*}[!htb]
    \includegraphics{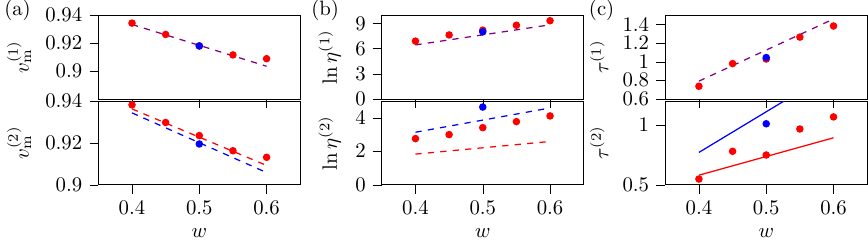}
    \caption{Evolution of fitting parameters in Eq.~\eqref{eq:ideal_ternary_SI}, corresponding to limit binary solutions $(1)$ (top row) and $(2)$ (bottom row), as a function of total PEG mass fraction $w$ and for mixtures of PEG-20000 $(1)$ with another PEG $(2)$ ({\color{red} $\bullet$} PEG-400, {\color{blue} $\bullet$} PEG-2000).  Dashed line corresponds to properties of binary solutions of water and PEG. (a) Specific volumes $v_\text{m}$, in $\SI{}{\centi\meter\cubed\per\gram}$. (b) Logarithm of viscosity $\ln \eta$ in $\SI{}{\milli\pascal\second}$. (c) Lifetime $\tau$, in $\SI{}{\nano\second}$.}
    \label{fig:SI_parameters_ternary_20000_tot}
\end{figure*}

\newpage

\section{More detailed investigation of PEG-400/PEG-2000 mixture}

\begin{figure*}[!htb]
    \includegraphics{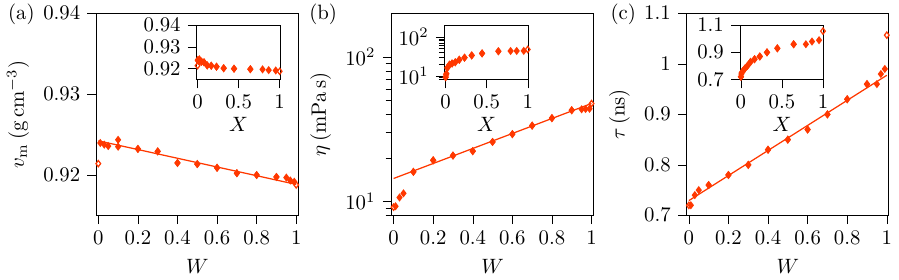}
    \caption{Properties of ternary mixture of water, PEG-400 and PEG-2000 at a total PEG mass fraction $w=0.5$, as a function of the mass proportion $W$ of PEG-2000. The open symbols on axis $W=0$ and $W=1$ represent data obtained in binary PEG/water mixtures. Continuous lines are linear fit on the range $W \in [0.1;0.9]$ as presented in main text. Inset represent the same data as a function of the mole proportion $X$ of PEG-2000. (a) Evolution of viscosity $\eta$ in semilogarithmic coordinates. (b) Evolution of specific volume $v_\text{m}$. (c) Evolution of lifetime $\tau$.}
    \label{fig:SI_400_2000_detailed_tot}
\end{figure*}

As discussed in main text about Fig.~5, there is an apparent discontinuity of properties of ternary mixture when approaching the limit of binary mixtures. Complementary measurements have been performed for mixture of PEG-400 and PEG-2000 for mass proportion $W$ of PEG-2000 in ranges $[0;0.1]$ and $[0.9;1]$. Results are represented on Fig.~\ref{fig:SI_400_2000_detailed_tot}.

As suggested in main text, these results show that this apparent discontinuity is actually hiding a progressive deviation of the apparent ideality upon approaching limit of binary solutions. This continuity is also well visible when plotting results as a function of mole proportion $X$ instead of mass fraction $W$. As highlighted in the main text, this shows that the mixing rules that are proposed are only valid in the range of composition considered experimentally.

\section{Details on the average lifetime measured in the local approach}

As discussed in main text, in the local approach, it can be considered that the rotor can be in contact with two possible environments, corresponding to limit solutions $(1)$ and $(2)$, and for which the lifetimes are respectively $\tau^{(1)}$ and $\tau^{(2)}$. As the collected light in FLIM is emitted by a macroscopic number of rotors present in the volume associated with each pixel and measured over acquisition time of a few second, significantly larger than typical time of molecular motion, it can be assumed that the obtained fluorescence lifetime is a superposition of the two lifetimes, with relative contribution given by the volume proportion $\Phi_1$ and $\Phi_2$ of the two environments.

In this picture, the lifetime distribution should be dual: this could be seen by using the polar plot representation of the FLIM~\cite{leray_2012}. In practice, the result is ambiguous as measurements do not fall exactly on the unit circle (which should be the case for monoexponential decays) but remains close. This is not contradictory as the two lifetimes $\tau^{(1)}$ and $\tau^{(2)}$ are not widely different. In presence of dual lifetime distribution, FLIM analysis provides an averaged value along~\cite{lakowicz_1990}:
\begin{equation}
\langle \tau \rangle = \frac{\Phi_1 \frac{(\tau^{(1)})^2}{1+ (\omega \tau^{(1)})^2} + \Phi_2 \frac{(\tau^{(2)})^2}{1+ (\omega \tau^{(2)})^2}}{\Phi_1 \frac{\tau^{(1)}}{1+ (\omega \tau^{(1)})^2} + \Phi_2 \frac{\tau^{(2)}}{1+ (\omega \tau^{(2)})^2}} 
\end{equation}
\noindent in which $\omega = 2\pi f$ is the pulsation of modulation of excitation light. In the considered experiments, $f=\SI{40}{\mega\hertz}$ and $\tau^{(1/2)} \sim \SIrange{0.1}{2}{\nano\second}$ so $(\omega \tau^{(1/2)})^2 \sim \SIrange{6e-3}{2e-1} \ll 1$ so the average lifetime is approximately given by:
\begin{equation}
\langle \tau \rangle = \frac{\Phi_1 (\tau^{(1)})^2 + \Phi_2 (\tau^{(2)})^2}{\Phi_1 \tau^{(1)} + \Phi_2 \tau^{(2)}}.
\end{equation}
\noindent Also, the reported lifetimes verify $\tau^{(2)} / \tau^{(1)} \sim \SIrange{1}{1.8}{}$. It can then be numerically verified that the following approximation is valid within a less than $\SI{10}{\percent}$ uncertainty (which anyway correspond to the typical uncertainty on measured lifetime):
\begin{equation}
\langle \tau \rangle = \Phi_1 \tau^{(1)} + \Phi_2 \tau^{(2)}.
\end{equation}

Finally, the volume fraction are defined as:
\begin{equation}
\Phi_i = \frac{V_i}{V_1 + V_2}
\end{equation}
\noindent in which $V_1$ and $V_2$ are the volumes occupied by the limit solutions $(1)$ and $(2)$ in the mixture. As their respective mass proportion are $W_i$, volume fraction is thus given by:
\begin{equation}
\Phi_i = \frac{W_i v^{(i)}}{W_1 v^{(1)} + W_2 v^{(2)}}.
\end{equation}
\noindent As the specific volumes $v^{(1)}$ and $v^{(2)}$ are in practice very close to each other, volume fractions $\Phi_i$ are approximately equal to mass fractions $W_i$ within a few percent. Lifetime returned by the FLIM analysis is thus given by:
\begin{equation}
\langle \tau \rangle = W_1 \tau^{(1)} + W_2 \tau^{(2)},
\end{equation}
\noindent which corresponds to the experimental observation reported in Fig.~5 of main text.

\end{document}